\documentclass[showpacs,twocolumn,aps,preprintnumbers,letterpaper]{revtex4}
\usepackage{amsmath,amssymb}
\usepackage{epsfig}
\usepackage{graphicx}
\usepackage{amsmath}
\usepackage{slashed}
\usepackage{amsfonts}
\usepackage{epstopdf}
\usepackage{color}

\newcommand{\be}{\begin{equation}}
\newcommand{\ee}{\end{equation}}
\newcommand{\bear}{\begin{eqnarray}}
\newcommand{\eear}{\end{eqnarray}}
\newcommand{\ba}{\begin{array}}
\newcommand{\ea}{\end{array}}

\def\be{\begin{eqnarray}}
\def\ee{\end{eqnarray}}
\def\bea{\be}
\def\eea{\ee}

\def\del{\partial}

\def\roughly#1{\mathrel{\raise.3ex\hbox{$#1$\kern-.75em%
\lower1ex\hbox{$\sim$}}}}

\begin{document}

\title{Hydrodynamics of the Chiral Dirac Spectrum}

\author{Yizhuang Liu}
\email{yizhuang.liu@stonybrook.edu}
\affiliation{Department of Physics and Astronomy, Stony Brook University, Stony Brook, New York 11794-3800, USA}

\author{Piotr Warcho\l{}}
\email{piotr.warchol@uj.edu.pl}
\affiliation{M. Smoluchowski Institute of Physics, Jagiellonian University, PL-30348 Krakow, Poland}

\author{Ismail Zahed}
\email{ismail.zahed@stonybrook.edu}
\affiliation{{}Department of Physics and Astronomy, Stony Brook University, Stony Brook, New York 11794-3800, USA}


\date{\today}
\begin{abstract}
We  derive a hydrodynamical description of the eigenvalues of the chiral Dirac spectrum in the vacuum and in the large 
$N$ (volume) limit. The linearized hydrodynamics supports sound waves. The stochastic relaxation of the eigenvalues is 
captured by a hydrodynamical instanton configuration which follows from a pertinent form of Euler equation. The 
relaxation from  a phase of localized eigenvalues and  unbroken chiral symmetry to a phase of de-localized  eigenvalues
and  broken chiral  symmetry occurs over a time set by the speed of sound. 
We show that the time is $\Delta \tau=\pi\rho(0)/2\beta N$ with $\rho(0)$ the spectral density at zero virtuality and $\beta=1,2,4$
 for the three Dyson ensembles that characterize QCD with different quark representations in the ergodic regime.
 \end{abstract}






\maketitle

\setcounter{footnote}{0}



{\bf 1. Introduction.\,\,} 
QCD with light quarks breaks spontaneously chiral symmetry. As a result, the light quarks transmute
to massive constituents which make up for most of the visible mass in our universe.  Empirical
evidence for the spontaneous breaking of chiral symmetry is in the form of light pions and kaons 
in nature~\cite{BOOK}. Dedicated  first principle QCD lattice simulations with light quarks have established
that chiral symmetry is spontaneously broken with a finite chiral condensate and an octet
of light mesons~\cite{LATTICE}.

The spontaneous breaking of chiral symmetry is characterized by a large accumulation of 
eigenvalues of the Euclidean Dirac operator near zero virtuality~\cite{CASHER}.  This phenomenon signals the
onset of an ergodic regime in the chiral Dirac spectrum. Other regimes where the light quarks 
diffuse or undergo ballistic motion can also be identified~\cite{DISORDER}. In many ways light quarks interacting
via colored Yang-Mills fields behave like disordered electrons in metallic grains. 

The essentials of the ergodic regime are captured by a chiral random matrix model~\cite{SHURYAK}. In short,
the QCD Dirac spectrum near zero virtuality  only retains that the QCD Dirac operator is chiral
or block off-diagonal, with random entries that are sampled from the three universal Dyson ensembles
~\cite{WAY} with a Gaussian weight thanks to  the central limit theorem. The spectrum corrections are also generic and
follow from the neighboring diffusive regime with the light quark return probability falling
like a power law in proper time~\cite{DISORDER}. Both regimes are separated by the Thouless energy
~\cite{DISORDER,OSBORN}.

In this letter we develop a hydrodynamical description of the eigenvalues of the QCD Dirac 
operator in the diffusive regime~\cite{DISORDER}, much along the lines of our recent studies of the 
eigenvalues of the Polyakov line at large number of colors~\cite{LIUPOL}. We will use this derivation to obtain the following new results:
1/ A hydrostatic solution for the spectral density beyond Wigner semi-circular distribution;
2/ a hydrodynamical instanton  that captures the stochastic relaxation of the eigenvalues of the
QCD Dirac operator for low virtuality; 3/ a dynamical relaxation time for restoring 
and/or breaking spontaneously chiral symmetry directly from the QCD Dirac spectrum;
4/ an estimate of this time for the dyon liquid model.

\vskip 0.5cm
{\bf 2. Chiral Dirac spectrum.\,\,}
The original chiral random matrix model partition function for the eigenvalues of the 
chiral Dirac spectrum was initially suggested as a null dynamical assumption for the generic 
analysis of the macroscopic chiral moments in the instanton liquid model~\cite{MACRO}. 
However and more importantly, it was noted that this assumption provides a universal description
of the chiral moments in the microscopic limit~\cite{SHURYAK,MICRO}. For QCD with $N_f$ quarks 
of equal masses $m$ in the complex representation~\cite{MACRO,SHURYAK}

\bea
Z_{2}[ m]=\int dT\, \prod_{f=1}^{N_f}{\rm det} \left(m^2+T^\dagger T\right)
\,e^{-\frac a2 N{\rm Tr}(T^\dagger T)}
\label{1}
\eea
Here $T$ is a symmetric random $C^{N\times N}$ complex matrix capturing the
random hopping between N-left and N-right zero modes. (\ref{1}) was generalized
to all Dyson ensembles with $\beta=1,2, 4$ corresponding to quarks in
different representations in~\cite{WAY},

\bea
Z_\beta[m]=
\int \prod^N_{i<j}|\lambda^2_i-\lambda^2_j|^{\beta}\prod^{N}_{i}
\lambda_i^{\alpha}(\lambda_i^2+m^2)^{N_f}e^{-\frac{a\beta N}{2}\lambda_i^2}d\lambda_i
\label{2}
\eea
with $\alpha=\beta(\nu+1)-1$. $\nu$ measures the topological asymmetry of $T$,
that is the difference between the number of zero modes and anti-zero modes 
for instance. Here $\lambda_i^2$ are  the eigenvalues of the squared Wishart matrix 
$W=T^\dagger T$. The overall normalization in (\ref{1}-\ref{2}) are omitted. 

The Gaussian 
measure is generic at large $N$ by the central limit theorem, with the
parameter $a$ fixed by the chiral condensate. In the microscopic limit
whereby the Dirac spectrum near zero-virtuality 
is magnified so that $N\lambda_i=1$, the interactions between the
eigenvalues as mediated by the gauge-fields are chaotic. As a result
only the chiral structure of the Dirac Hamiltonian and the symmetry of the 
matrix entries under time-reversal are relevant~\cite{ALL}. This is the universal
regime of ergodicity shared by most disordered electronic systems
in the mesoscopic limit~\cite{MESO}.

Alternatively, (\ref{2}) can be regarded as the normalization of the squared and
real many-body wave-function 

\be
Z_\beta[m]=\int \prod_{i=1}^Nd\lambda_i\,\left|{\bf \Psi}_0[\lambda_i]\right|^2
\label{2X}
\ee
which  is the zero-mode solution to the Shrodinger equation $H_0\Psi_0=0$ with
the self-adjoint squared Hamiltonian

\be
H_0\equiv\sum_{i=1}^{N}\left(-\partial_i+{\bf a}_i\right)\left(\partial_i+{\bf a}_i\right)
\label{3}
\ee
with $\partial_i\equiv \partial/\partial\lambda_i$ and the pure
gauge potential ${\bf a}_i\equiv \partial_iS$. 
The mass parameter is $1/2$.
The Vandermond contribution $\Delta=\prod_{i<j}|\lambda^2_{ij}|^{\beta}$ induces a diverging
2-body part. It is convenient to re-absorb it using a similarity transformation by defining 
$\Psi=\Psi_0/\sqrt{\Delta}$ and the new Hamitonian 

\be
H=(1/\sqrt{\Delta})H_0\sqrt{\Delta}
\label{3X}
\ee

\vskip 0.5cm
{\bf 3. Hydrodynamics.\,\,}
We can use the collective coordinate method in~\cite{JEVICKI}  to re-write (\ref{3X}) in terms of the
density of eigenvalues as a collective variable $\rho(\lambda) =\sum^{N}_{i=1}\delta(\lambda-\lambda_i)$. 
After some algebra we obtain

\be
H=\int d\lambda \left(\partial_{\lambda}\pi\rho\, \partial_{\lambda}\pi+ \rho {\bf u}[\rho]\right)
\label{4}
\ee
with the potential-like contribution ${\bf u}[\rho]\equiv{\bf A}^2$

\bea
{\bf A}
=&&\frac{a\beta N}{2}\lambda-\frac{\beta\pi}{2}[\rho_H(\lambda)-\rho_H(-\lambda)]\nonumber\nonumber\\
&&+\partial_\lambda{\rm ln}\sqrt{\rho}-\frac{\alpha}{2\lambda}-\frac{N_f\lambda}{\lambda^2+m^2}
\label{5}
\eea
Here  $\rho_H$ is the Hilbert transform of $\rho$

\be
[\rho]_H\equiv \rho_H(\lambda)=\frac{1}{\pi}{\rm P}\int d\lambda^\prime \frac{\rho(\lambda^\prime)}{\lambda-\lambda^\prime}
\label{6}
\ee
The canonical pair, $\pi(\lambda)$ and  $\rho(\lambda)$  satisfies
the Poisson bracket
$\{\pi(\lambda),\rho(\lambda^\prime)\}=\delta(\lambda-\lambda^\prime)$.
We identify the collective fluid velocity with $v=\partial_\lambda\pi$
and re-write (\ref{4}) in the more familiar hydrodynamical form

\be
H\approx \int d\lambda \rho(\lambda) \left( v^2+ {\bf u}[\rho]\right)\approx \int d\theta \rho(\lambda) \left|v+i{\bf A}\right|^2
\label{7X1}
\ee
ignoring ultra-local terms. 
The equation of motion for $\rho$ yields the current conservation law 
$\partial_t\rho=-2\partial_\lambda \left(\rho  v\right)$, and the equation 
of motion for $v$ gives the  Euler equation

\bea
\label{10} \nonumber
&&\partial_t v=\{H, v\}=
-\partial_\lambda\left[v^2+{\bf A}^2-\partial_\lambda {\bf A}
-{\bf A}{\partial_\lambda{\rm ln}\rho}\right.\\
&&\left.+\pi \beta[{\bf A}\rho]_H (\lambda) +\pi \beta[{\bf A}\rho]_H (-\lambda)\right]
\eea
All the relations  hold for large but finite $N$, to allow for a smoothening of the density.

\vskip .5cm
{\bf 4. Hydrostatic solution.\,\,}
The hydrostatic solution corresponds to the minimum of (\ref{7X1}) with $v=0$
and thus ${\bf A}(\lambda)=0$.  In terms of the chirally symmetric combination
 $\rho^\chi(\lambda)\equiv \rho(\lambda)+\rho(-\lambda)$  we have
 
\bea
0
=&&\frac{a\beta N}{2}\lambda-\frac{\beta\pi}{2}\rho^\chi_H(\lambda)\nonumber\nonumber\\
&&+\frac 12 \partial_\lambda{\rm ln}\rho^\chi-\frac{\alpha}{2\lambda}-\frac{N_f\lambda}{\lambda^2+m^2}
\label{rhochistat}
\eea
In the large $N$ limit only the terms in the first line survive with

\be 
{\rho_0^\chi}(\lambda)=\frac {Na} \pi \sqrt{\frac 4a -\lambda^2}
\label{RHO}
\ee
which is Wigner Semi-circular distribution within the support $|\lambda|\leq 2/\sqrt{a}$. To correct it in $1/N$ we define
$\rho^\chi\approx \rho_0^\chi+\rho_1$  subject to the condition
$\int \rho_1 {\rm d}\lambda =0$,  so that

\be  
-\beta \pi {\rho_1}_H(\lambda)=\frac {1}{\lambda}(\alpha +2N_f)-\frac{\lambda a}{4-a\lambda^2}
\ee 
The closed form solution is

\bea
\label{RHO1}
\rho_1(\lambda)=&&
-\frac{\alpha+2N_f}{\beta}\,\delta(\lambda)\nonumber +\frac{|\lambda|}{\beta\sqrt{4/a}}
\,\delta(\lambda^2-{4/a})\nonumber\\
&&+\frac{\alpha+2N_f-\frac{1}{2\sqrt{4/a}}}{\beta\pi\sqrt{4/a -\lambda^2}}
\eea
with the support unchanged. We note that (\ref{RHO1}) shows a delta-function accumulation
at zero virtuality with the negative strength $\alpha+2N_f$.

The general solution to (\ref{rhochistat})  for large but finite $N$ can be sought in the massless case or $m=0$,
by multiplying (\ref{rhochistat}) by $2\rho^\chi \lambda (\lambda^2+m^2)$ and taking 
the Hilbert transform. The result is

\be \nonumber
a\beta N \left(\lambda^2 \rho_H^\chi-\lambda \frac {2N}{\pi}\right) -\frac{\beta \pi}{2} \lambda\left({\rho_H^\chi}^2 - {\rho^\chi}^2\right)\\ +\lambda \del_\lambda \rho_H^\chi -(\alpha +2N_f)\rho_H^\chi =0
\label{XL1}
\ee  
To solve it consider the extension of the resolvent 
\be
G(z)=\int d\lambda \frac{\rho(\lambda)}{z-\lambda}
\ee
to the upper and lower complex plane,

\be
G^{\pm}(z\to\lambda)=\pi\rho_H^\chi (\lambda) \mp i\pi \rho^\chi (\lambda)
\label{XL2}
\ee
so that (\ref{XL1}) now reads

\be 
a\beta N z^2 G -\frac 12 \beta z G^2 +z\del_z G-(\alpha +2N_f)G -2a\beta N^2 z=0\nonumber\\
\label{XL3}
\ee
Defining $G(z)\equiv -(2/\beta){\del_z {\rm ln}f(z)}$ and setting
$z^2=-w$ and $f(\sqrt{-w})=g(w)$, we have

\be  
\label{XL4}
2w\del_{ww}g +\left(- a\beta N w +1 -\alpha -2 N_f\right)\del_w g - \frac 12 a \beta^2 N^2 g =0\nonumber\\
\ee
We note that (\ref{XL4}) is Laguerre-like except for the wrong sign in the last contribution. 
The general solution is a hyper-geometric function. The spectral density follows from the discontinuities
of the solution using (\ref{XL2}).

\vskip .5cm
{\bf 5. Dyson Coulomb gas.\,\,}
We note that (\ref{2}) can be re-written in terms of 1-dimensional Dyson Coulomb
gas.  At large $N$ the ensemble described by (\ref{1}) is sufficiently dense  to allow the change in the measure,

\be
\prod^N_{i=1}d\lambda_i\approx e^{{\cal S}[\rho ]} D\rho
\label{11}
\ee
with ${\cal S}[\rho]=\int d\lambda\rho(\lambda){\rm ln}(\rho_0/\rho(\lambda))$ the Boltzmann entropy~\cite{MEHTA}.
Thus

\be
Z_\beta[m]\rightarrow \int D\rho \,e^{-\Gamma[\beta, m ; \rho]}
\label{12}
\ee
with the effective action

\bea
&&\Gamma[\beta, m ; \rho]=\nonumber\\
&&+\int d\lambda\rho(\lambda)\left(\frac{a\beta N}{2}\lambda^2-\alpha{\rm ln}(\lambda)-N_f{\rm ln}(\lambda^2+m^2)\right)
\nonumber \\
&& -\frac{\beta}{2}\int d\lambda d\lambda' \rho(\lambda)\rho(\lambda'){\rm ln}(\lambda^2-\lambda'^2)\nonumber\\
&&-\left(\frac\beta 2- 1\right)\int d\lambda \rho {\rm  ln} \rho
\label{13}
\ee
The $\beta$ contribution is the self Coulomb subtraction and is consistent with the subtraction in the Hilbert transform.
The saddle point equation $\delta\Gamma/\delta\rho=0$ yields the hydro-static equation (\ref{rhochistat}) using the symmetric density
$\rho^\chi$.  


\vskip 0.5cm
{\bf 6. Hydrodynamical  instanton.\,\,} Following the initial observation in~\cite{LIUPOL}, we note
that the  fixed time  zero energy solution to (\ref{7X1}) is an instanton with imaginary velocity
$v=-i {\bf A}$. The conserved current $j\equiv \rho v$
satisfies  ($\tau=it$)

\bea
\partial_\tau\rho(\lambda) = 
\del_\lambda\left[a\beta N \lambda \rho(\lambda)-\beta\pi \rho(\lambda)[\rho_H(\lambda)-\rho_H(-\lambda)]\right.\nonumber\\
\left.+\partial_\lambda \rho(\lambda)-\frac{\alpha}{\lambda}\rho(\lambda)-\frac{2 N_f\lambda}{\lambda^2+m^2}\rho(\lambda) \right]
\label{15}
\eea
 We identify $\tau$ with the stochastic  time, and 
(\ref{15}) describes the stochastic relaxation of the  chiral eigenvalue density of the 
chiral Dirac spectrum (out of equilibrium)  to its asymptotic (in equilibrium) hydro-static solution. 
For $m=0$, (\ref{15}) can be rewritten as

\be \nonumber
\del_\tau \rho^\chi=a\beta N\del_\lambda \left(\lambda \rho^\chi\right)-\beta\pi \del_\lambda \left(\rho^\chi \rho_H^\chi\right)\\
+\del_{\lambda\lambda}\rho^\chi -(\alpha +2N_f)\del_\lambda \left(\frac 1\lambda \rho^\chi\right)
\ee
After multiplying this equation by $\lambda^2$, we can take its Hilbert transform. The result is
\be \nonumber
\del_\tau \rho_H^\chi=a\beta N\del_\lambda \left(\lambda \rho_H^\chi\right)-\frac{\beta\pi}{2}\del_\lambda \left({\rho_H^\chi}^2 -{\rho^\chi}^2\right)\\
+\del_{\lambda\lambda}\rho_H^\chi -(\alpha +2N_f)\del_\lambda \left(\frac 1\lambda \rho_H^\chi\right)
\ee
With the usual definition, this gives an equation for the time dependent Green's function
\be \nonumber
\del_\tau G=a\beta N\del_z \left(z G\right)-\frac{\beta}{2} \del_z \left(G^2\right)\\
+\del_{zz}G -(\alpha +2N_f)\del_z \left(\frac 1z G\right)
\ee
We note that the stationary solution fulfills

\be 
a\beta N z G-\frac{\beta}{2} G^2
+\del_{z}G -(\alpha +2N_f)\frac 1z G+C=0
\label{X01}
\ee
The constant $C$ is fixed by noting that for large $N$ and large $z$, $G/N\approx 2/z$ so that $C=4aN^2$,
in agreement  with (\ref{XL3}).

The general time-dependent solution can be analyzed by first re-scaling $\tau\to\tau /N$. In the large $N$ limit $G$ solves

\be
\del_\tau G+\beta\left(\frac{1}{N}G-az\right)\del_z G - a\beta G=0
\ee
which is a Burgers like nonlinear PDE.  We can solve it with the method of complex characteristics~\cite{CHBNW}. 
For that we introduce curves in the  space of $z$ and $\tau$, parametrized by $s$ and labeled by $z_0$, 
along which the nonlinear PDE follows from ordinary ODE

\bea
&&\frac{{\rm d} z}{{\rm d}s}=\beta\left(\frac{1}{N}G-az\right)\nonumber\\
&&\frac{{\rm d} G}{{\rm d}s}=a\beta G\nonumber\\
&&\frac{{\rm d} \tau}{{\rm d}s}=1
\label{X02}
\eea
with the condition that $z(\tau=0)=z_0$ and $\tau(s=0)=0$ (which means $\tau=s$). These ODE
can be solved for a specific initial condition.

\vskip 0.5cm
{\bf 7. Sound waves.\,\,}
To gain some insights to the general time-dependent solutions, we analyze first 
the hydrodynamical equations in the linearized density approximation. For that, it is
convenient to identify the classical hydrodynamical action associated to
(\ref{4}). Using the standard canonical procedure we found 

\be
{\bf S}=\int dt d\lambda\,\rho(\lambda)\left(v^2-{\bf u}[\rho]\right)
\ee
which is linearized by

\bea
\rho\approx\rho_0(\lambda)+2\partial_{\lambda}\varphi \qquad{\rm and}\qquad
\rho v\approx- \partial_t\varphi
\label{16}
\eea
Inserting (\ref{16}) into ${\bf S}$ yields in the quadratic approximation

\be
{\bf S}_2=\int dt \frac {d\lambda}{\rho_0(\lambda)}  \, 
\left({(\partial_t \varphi)^2}-{\rho^2_0(\lambda)}{W}^2[\varphi]\right)
\label{17}
\ee
with the potential

\be
W[\varphi]=\beta \pi \partial_\lambda\left[\varphi_H(\lambda)+\varphi_H(-\lambda)\right]-
\partial_\lambda\left(\frac{\partial_\lambda{\varphi}}{\rho_0(\lambda)}\right)
\label{18}
\ee
For  $\rho(\lambda)=\rho_0(\lambda)$, after the rescaling $Nt\rightarrow t$, (\ref{17}) simplifies for large $N$

\bea
\label{19}
&&{\bf S}_2\approx  N^2
\int dt \frac {d\lambda}{\rho_0(\lambda)} \\
&&\times
\left({(\partial_t \varphi)^2}-\frac{(\pi\beta \rho_0(\lambda))^2}{N^2}\left(\partial_{\lambda}\varphi_H(\lambda)+\partial_{\lambda}\varphi_H(-\lambda)\right)^2\right)\nonumber
\eea
The speed of sound $v_s(\lambda)=2\pi\beta\rho_0(\lambda)/N$ is local in the chiral spectrum.
It is $v_s(0)=2\beta\sqrt{a}$ at zero virtuality.
We note that (\ref{19}) is extensive with $N$ for $\rho_0(\lambda)/N$ normalized to 1.

\vskip 0.5cm
{\bf 8. Chiral relaxation time.\,\,}An interesting dynamical question regarding the chiral Dirac spectrum 
is the typical relaxation time associated to the formation or disappearance of the spontaneous breaking of
chiral symmetry. This issue is important  for QCD with matter undergoing dynamical
chiral symmetry restoration or breaking close to the  chiral transition temperature. Since the set-up 
involves off-equilibrium QCD in a moderately strong coupling regime, first principle calculations
are usually challenging.

Here we answer this question  by focusing 
on the time it takes for the eigenvalues near zero-virtuality viewed as
a hydrodynamical fluid to re-arrange stochastically. For that, consider that at 
$\tau=0$ all the eigenvalues are localized at zero virtuality to mock up 
an initial and chirally restored phase. The relaxation of this
phase to a spontaneously broken chiral phase is characterized by the
time it takes the sound wave to fill up the gap or the so-called zero-mode-zone (ZMZ)
spanned by Wigner semi-circle.

To show this we use the initial condition $G(z=z_0, \tau=0)={2N}/{z_0}$ 
in (\ref{X02}) to obtain the solution

\be \label{gtau}
G=e^{a\beta\tau} \frac{2N}{z_0}
\ee 
The remaining equation is now

\be 
\frac{{\rm d} z}{{\rm d}\tau}=\beta\left(\frac{2}{z_0}e^{a\beta\tau}-a z\right)
\ee 
which yields
\be 
z= \frac 1{az_0} e^{a\beta\tau}+\left(z_0 -\frac 1 {az_0}\right)e^{-a\beta\tau}
\ee 
Solving for $z_0$ and inserting the answer in (\ref{gtau}) yield

\bea 
\frac {G(z)}{N\sqrt a}=&&\left(1-e^{-v_s(0)\tau\sqrt{a}}\right)^{-1}\\
&&\times  \left(z\sqrt{a} \pm \sqrt{(z\sqrt a)^2 -4\left(1-e^{-v_s(0)\tau\sqrt{a}}\right)}\right)\nonumber
\eea 
The spectral density follows from the discontinuity of (\ref{RHO})

\bea
\rho(\lambda)=&&\frac{aN}{\pi}
\left(1-e^{-v_s(0)\tau\sqrt{a}}\right)^{-1}\\
&&\times  \left(\frac{4}{a}\left(1-e^{-v_s(0)\tau\sqrt{a}}\right)-\lambda^2\right)^{\frac 12}\nonumber
\eea
which is seen to interpolate between a delta-function $N\delta(\lambda)$ at $\tau=0$ and a Wigner semi-circle at asymptotic $\tau$
at a rate given by the speed of sound $v_s(0)$ at zero virtuality. The typical time in physical units is 

\be
\Delta \tau\equiv\frac a{v_s(0)}=\frac{\pi \rho(0)}{2\beta N}=\frac{|\left<q^\dagger q\right>|}{2\beta{\bf n}}
\ee
Recall that in chiral random matrix theory the scale $a$ is related to the chiral condensate by
the Banks-Casher formula $V_4\left<q^\dagger q\right>=-\pi\rho(0)=- \,N\sqrt{a}$. Thus, the last
equality with ${\bf n}=N/V_4$. Near zero virtuality and close to the ergodic
regime this time  is universal. It is non-perturbative and fully gauge-invariant. 

In the QCD vacuum we may identify ${\bf n}$ with the instanton density~\cite{BOOK} (and references therein).
Typically,  $|\left<q^\dagger q\right>|\approx 1/(1\,{\rm fm}^3)$ and ${\bf n}\approx 1/(1\,{\rm fm}^4)$ 
 so that $\Delta t\approx 1\,{\rm fm}/4$ for QCD with $N_c=3$ ($\beta=2$),
 which is short. In matter, both the chiral condensate
 and the instanton density change. For a dyon liquid model in the confining phase
 in the range $0.5<T/T_c<1$,  we may identify ${\bf n}={\bf n}_D/N_c$ with ${\bf n}_D$ the dyon density~\cite{DYON}. 
 We recall that the instanton splits to $N_c$ dyons with only the KK dyon carrying the zero-mode.
 From the analysis presented in~\cite{LIUFERMION} we have for $N_c=2$ ($\beta=1$) and $N_f=1$

\be
\Delta \tau=\frac{|\left<q^\dagger q\right>|}{2\beta{\bf n_D}/N_c}
\approx \frac{1.25}{T}\left(C\frac{e^{-\frac \pi{\alpha_s(T)}}}{\alpha_s^2(T)}\right)^{0.63}
\ee
with  $\alpha_s(T)=(10/3)\,{\rm ln}({T}/{0.36 T_c})$ the running coupling, and
$C$ a constant of order 1 in the dyon density which is fixed by lattice measurements.



\
\vskip 0.5cm
{\bf 9. Conclusions.\,\,}The hydrodynamical description of the chiral Dirac spectrum captures some key aspects of the
stochastic relaxation of the Dirac eigenvalues in the diffusive regime. The hydrodynamical set-up supports an
instanton  that describes  the stochastic relaxation of the Dirac eigenvalues as a fluid. The fluid exhibits sound waves
that can be used to estimate the time it takes for a localized density with chiral symmetry restored, to de-localize to
a Wigner semi-circle with chiral symmetry restored.  The relaxation time of a fluid 
of Dirac eigenvalues near zero virtuality captures a  typical equilibration time for chiral symmetry breaking or restoration
in QCD that is robust and gauge-independent.

\vskip 0.25cm
{\bf Acknowledgements}
The work of YL and IZ is  supported in part  by the U.S. Department of Energy under Contracts No.
DE-FG-88ER40388. The work of PW is supported by the DEC-2011/02/A/ST1/00119 grant and 
the UMO-2013/08/T/ST2/00105 ETIUDA scholarship of the (Polish) National Centre of Science. 
PW thanks Stony Brook University for its hospitality during the completion of this work.
 \vfil
\end{document}